\documentclass[lettersize,journal]{IEEEtran}
\usepackage{amsmath,amsfonts}
\usepackage{algorithmic}
\usepackage{algorithm}
\usepackage{array}
\usepackage{stfloats}
\usepackage{subfig} %

\usepackage{comment}
\usepackage{graphicx}
\usepackage{textcomp}
\usepackage{stfloats}
\usepackage{url}
\usepackage{verbatim}
\usepackage{graphicx}
\usepackage{cite}
\usepackage{orcidlink}

\usepackage{xcolor}
\usepackage{booktabs}
\definecolor{Mygreen}{rgb}{0.00, 0.72, 0.0}
\usepackage{authblk}

\begin{document}

\title{GroundBIRD Telescope: Systematics Modelization of MKID Arrays Response}

\author{Yonggil Jo*\,\orcidlink{0000-0002-4340-3171}, 
Alessandro Fasano\,\orcidlink{0000-0003-4041-418X},
Eunil Won\,\orcidlink{0000-0002-4245-7442},
Makoto Hattori\,\orcidlink{0000-0003-0620-2554},
Shunsuke Honda\,\orcidlink{0000-0002-0403-3729},
Chiko Otani\,\orcidlink{0000-0002-9406-2602},
Junya Suzuki\,\orcidlink{0000-0001-6816-8123}, 
Mike Peel\,\orcidlink{0000-0003-3412-2586},
Kenichi Karatsu\,\orcidlink{0000-0003-4562-5584},
Ricardo Génova-Santos\,\orcidlink{0000-0001-5479-0034},
and Miku Tsujii\,\orcidlink{0009-0000-6741-6033}

\thanks{Email: zdragonroadz@gmail.com\\
Yonggil Jo and Eunil Won are with the Department of Physics, Korea University, Seoul, South Korea.}%
\thanks{Alessandro Fasano and Ricardo Génova-Santos are with the Instituto de Astrofísica de Canarias, E-38200 La Laguna, Tenerife, Spain / Departamento de Astrofísica, Universidad de La Laguna, E-38206 La Laguna, Tenerife, Spain.}%
\thanks{Makoto Hattori and Miku Tsujii are with the Astronomical Institute, Tohoku University, 2 Chome-1-1 Katahira, Aoba Ward, Sendai, Miyagi 980-8577, Japan.}%
\thanks{Shunsuke Honda is with the University of Tsukuba, 1 Chome-1-1 Tennodai, Tsukuba, Ibaraki 305-8577, Japan.}%
\thanks{Chiko Otani is with Center for Advanced Photonics, RIKEN, 519-1399 Aramaki-Aoba, Aoba-ku, Sendai, Miyagi 980-0845, Japan / the Department of Physics, Tohoku University, 6-3 Aramaki-Aoba, Aoba-ku, Sendai 980-8578, Japan.}%
\thanks{Junya Suzuki is with the Division of Physics and Astronomy, Graduate School of Science, Yoshidahonmachi, Sakyo Ward, Kyoto, 606-8501, Japan.
}
\thanks{Mike Peel is with the Department of Astrophysics, Imperial College London, South Kensington Campus, London SW7 2AZ, United Kingdom.
}%
\thanks{Kenichi Karatsu is with SRON Netherlands Institute for Space Research, Niels Bohrweg 4, 2333 CA Leiden, Netherlands.
}%
}

\maketitle
\IEEEpubid{\begin{minipage}{\textwidth}
\vspace{25pt}
\footnotesize
\raggedright  
© 2026 IEEE. Personal use of this material is permitted. Permission from IEEE must be obtained for all other uses, in any current or future media, including reprinting/republishing this material for advertising or promotional purposes, creating new collective works, for resale or redistribution to servers or lists, or reuse of any copyrighted component of this work in other works.
This article has been accepted for publication in IEEE Transactions on
Applied Superconductivity. The final published version will be available upon publication via IEEE Xplore at DOI: 10.1109/TASC.2026.3672082.
\end{minipage}}

\begin{abstract}
Kinetic inductance detectors are widely used in millimeter- and submillimeter-wave astronomy, benefiting from their fast response and relative ease of fabrication. The GroundBIRD telescope employs microwave kinetic inductance detectors at 145 and 220 GHz to observe the cosmic microwave background. As a ground-based telescope, it is subject to inherent environmental systematics, namely atmospheric emission and thermal fluctuations of the focal plane temperature. This study models resonance frequency shifts induced by each source using calibrated on-site measurements of precipitable water vapor and temperature. Comparison with observational data confirms the validity of the models and identifies atmospheric loading as the dominant contributor to frequency variation under typical observation conditions.
\end{abstract}


\section{Introduction}
\IEEEpubidadjcol
\IEEEPARstart{L}{arge}-angular scale measurements of the cosmic microwave background (CMB) with higher sensitivity would improve constraints on the optical depth to reionization, which is the most weakly constrained parameter in the standard Lambda-Cold Dark Matter ($\Lambda$CDM) model \cite{Planck}\cite{Frieman2008}. Furthermore, the large-scale polarization signal of the CMB offers a means to detect primordial gravity waves \cite{zaldarriaga} \cite{Chen2025}.
Achieving these scientific goals requires not only highly sensitive detector arrays with low noise and rapid response, but also a deep understanding of environmental and instrumental systematics. 

\begin{figure}[t]
  \centering
    \includegraphics[width=0.95\columnwidth]{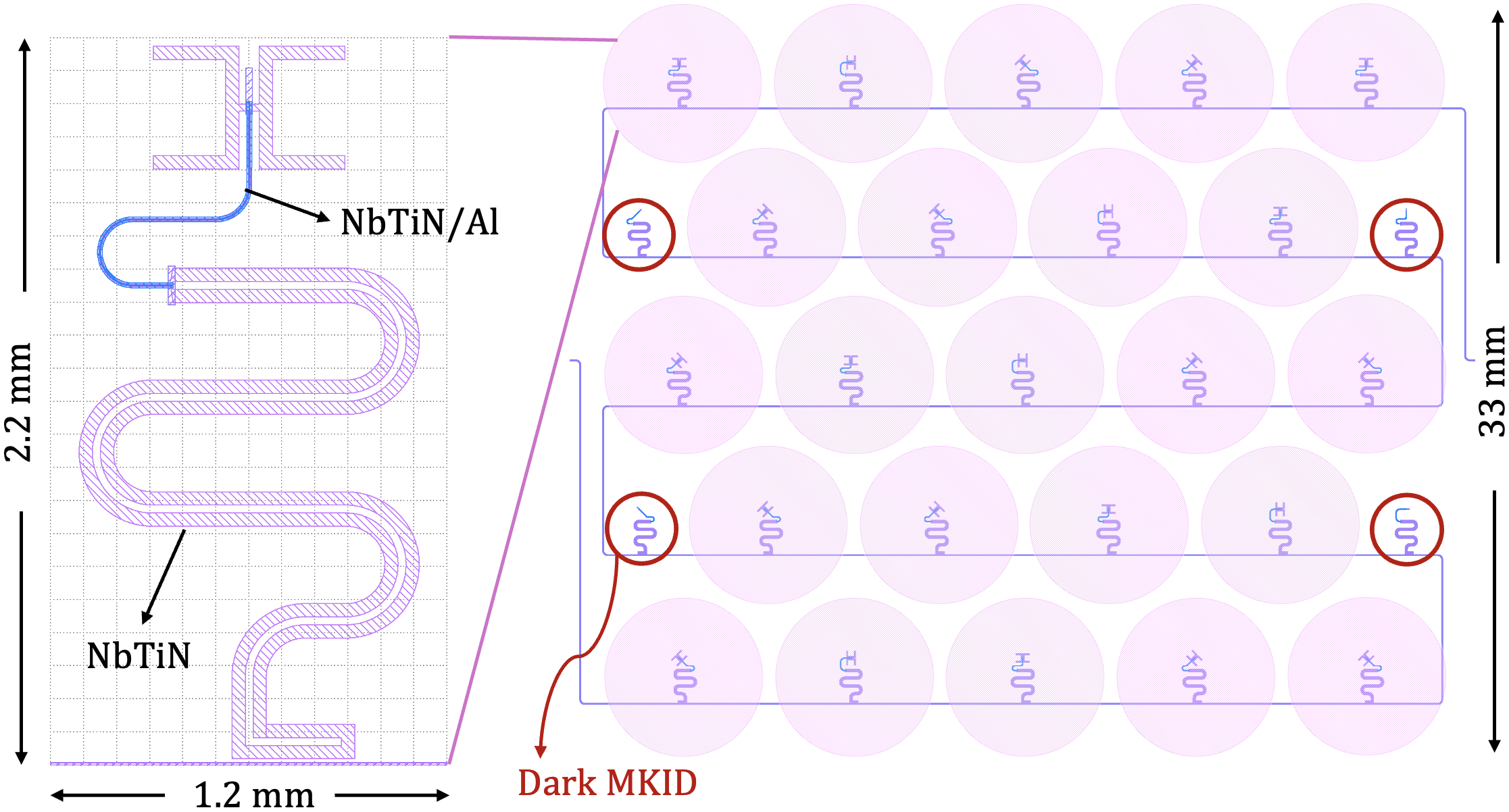}
  \vspace{0.5em}
  \caption{Schematic of a $145$ GHz MKID array installed in the GroundBIRD focal plane. The left panel depicts an individual MKID, with the active resonator section composed of NbTiN and Al, and the coupling region made solely of NbTiN. The right panel shows the full layout of an array containing $23$ MKIDs with lenses and antennas. The dark MKIDs are highlight by red circles.}
  \label{fig:chip schematic}
\end{figure}

The Ground-based B-mode polarization and Inflation from cosmic background Radiation Detection telescope (GroundBIRD telescope)\cite{tsuji-commissioning} employs NbTiN/Al hybrid Microwave Kinetic Inductance Detectors (MKIDs) \cite{Zmuidzinas} with hemispherical lenslets and twin-slot antenna coupled to detectors, operating at cryogenic temperatures. The focal plane of the GroundBIRD telescope consists of seven MKID arrays; one $220$ GHz array at the center surrounded by six $145$ GHz arrays at the edges. Each array contains $23$ illuminated MKIDs and four dark MKIDs without a lens and an antenna. Each array is assigned a unique id, and the results presented in this study are from four of the $145$ GHz arrays, $\textrm{2A}\textrm{, } \textrm{1B}\textrm{, }  \textrm{2B}\textrm{, }\textrm{and}\:\textrm{3B}$. Fig. \ref{fig:chip schematic} shows the layout schematic of a $145$ GHz array. Close-up of a single MKID and a coupled antenna is shown on the left, where the thin blue line shows the active part of the resonator composed of NbTiN and Al. The coupling part of the resonator is composed of NbTiN. The right figure shows the full layout of an array. Each purple-shaded circle represents a hemispherical lenslet composed of silicon, and red circles highlight the dark MKIDs dedicated to the analysis of non-optical systematics. Detailed  dimensions, layer stacks, and design processes are provided in \cite[pp.~29--41]{tomonaga-phd}.

\IEEEpubidadjcol
The telescope is located at the Teide Observatory in Tenerife, of the Canary Islands (Spain). The telescope scans the sky with a continuous azimuthal rotation speed of potentially up to $20$ rotations per minute (RPM), a fast-scanning strategy that mitigate $1/f$ noise \cite{Nagasaki2018} and achieves a sky coverage of approximately $40\%$. 

MKIDs operate by detecting changes in quasiparticle population within a superconducting resonator. Energy from incident photons or thermal fluctuations breaks Cooper pairs, increasing the kinetic inductance that shifts the resonance peak to a lower frequency \cite{Day2003} \cite{mazin-phd}.

Ground-based CMB telescopes, such as GroundBIRD are susceptible to variations in atmospheric emission, even at high-altitude observation sites \cite{Ananthasubramanian2002}. In particular, fluctuations in precipitable water vapor (PWV) affect atmospheric transparency and emissivity \cite{Radford_2016}, introducing time-varying optical loads on the detectors. In addition to atmospheric effects, thermal fluctuations inside the cryostat---partially driven by the azimuthal rotation of the telescope---introduce further systematics by altering the focal plane temperature.

In this paper, we present the characterizations of environmental systematics that affect resonance frequencies of the MKIDs in the GroundBIRD telescope, attributing the response to variations in quasiparticle density. First, we model the effects of atmospheric emission, parametrized by the calibrated on-site PWV measurements. Second, effects from the thermal fluctuations of the focal plane are modeled and tested using measurements from dark MKIDs.

\section{Observation Cycle and Calibration Strategy}
The GroundBIRD readout system uses the custom board named "Rhea" \cite[pp.~30--33]{yoshinori-phd}, a wide-bandwidth analog front-end for MKID arrays. It features 200 MHz digital-to-analog and analog-to-digital converters for intermediate frequency signal processing from $-100 \:\mathrm{MHz}$ to $+100 \:\mathrm{MHz}$. The local oscillator is tunable from $4\:\mathrm{GHz}$ to $8\:\mathrm{GHz}$ for the up-conversion.

To trace the evolution of the incoming signal, each observation cycle begins with a frequency sweep measurement to locate the resonance frequencies of all MKIDs. 
This procedure is repeated hourly. Each MKID's resonance frequency is estimated by fitting its peak in complex in-phase and quadrature (IQ) data from frequency sweep measurements to a slightly modified version of the transmission model shown in \cite{gao-phd} to include a linear correction term, given as
\begin{align}
\label{eq:lingao}
t_{21}(f)&=ae^{-2\pi if\tau_{c}} \left[ 1+c(f-f_{r})-\frac{Q_r/Q_ce^{i\phi_0}}{1+2iQ_r(\frac{f-f_r}{f_r})} \right]. 
\end{align}
The fit determines the cable delay constant $\tau_c$, the linear term constant $c$, the resonance frequency $f_r$, the total resonance quality factor $Q_r$, the coupling quality factor $Q_c$, and the phase offset $\phi_0$. The typcial value of $\langle Q_r\rangle=(1.4\pm0.3)\times 10^4$, and $\langle Q_c\rangle=(3\pm1)\times10^4$, measured under laboratory conditions \cite{tomonaga-mkid}.
 
\begin{figure}[t]
  \centering
    \includegraphics[width=0.99\columnwidth]{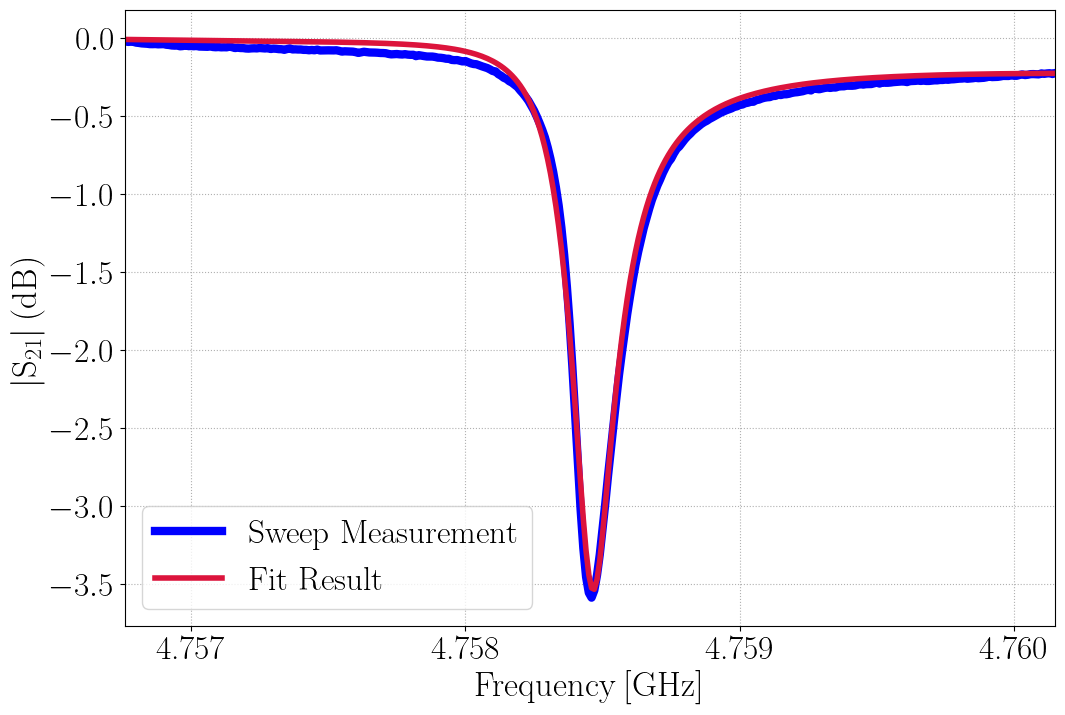}
  \caption{Frequency sweep measurement (blue) and its fit result to the transition model (red) of MKID $1$ on $\textrm{Array1B}$. Fitted parameters include resonance frequency, quality factors, cable delay, and phase offset.}
  \label{fig:peak sweep}
\end{figure}

The frequency sweep data of a single detector and its fit are shown in Fig. \ref{fig:peak sweep}. 
Each MKID also has an off-resonance readout, which is utilized to correct for gain variations and common-mode electronic systematics in the data processing pipeline \cite{tomonaga-phd}. After the resonance frequencies are estimated through the sweep measurement, the system transitions to fixed-frequency time-ordered-data (TOD) acquisition. The TOD records the phase evolution of each detector by continuously measuring the phase shifts in the IQ plane corresponding to the resonance frequency changes.

In addition to systematics arising from variable background signals such as atmospheric radiation, MKIDs---though known to have a linear frequency response \cite{Monfardini2014}---exhibit residual systematics when phase readout is employed. These effects stem from limitations in the current readout implementation, which does not fully track changes in the background environment.

\begin{figure*}[!t]
\centering
\subfloat[]{
\includegraphics[width=0.985\columnwidth]{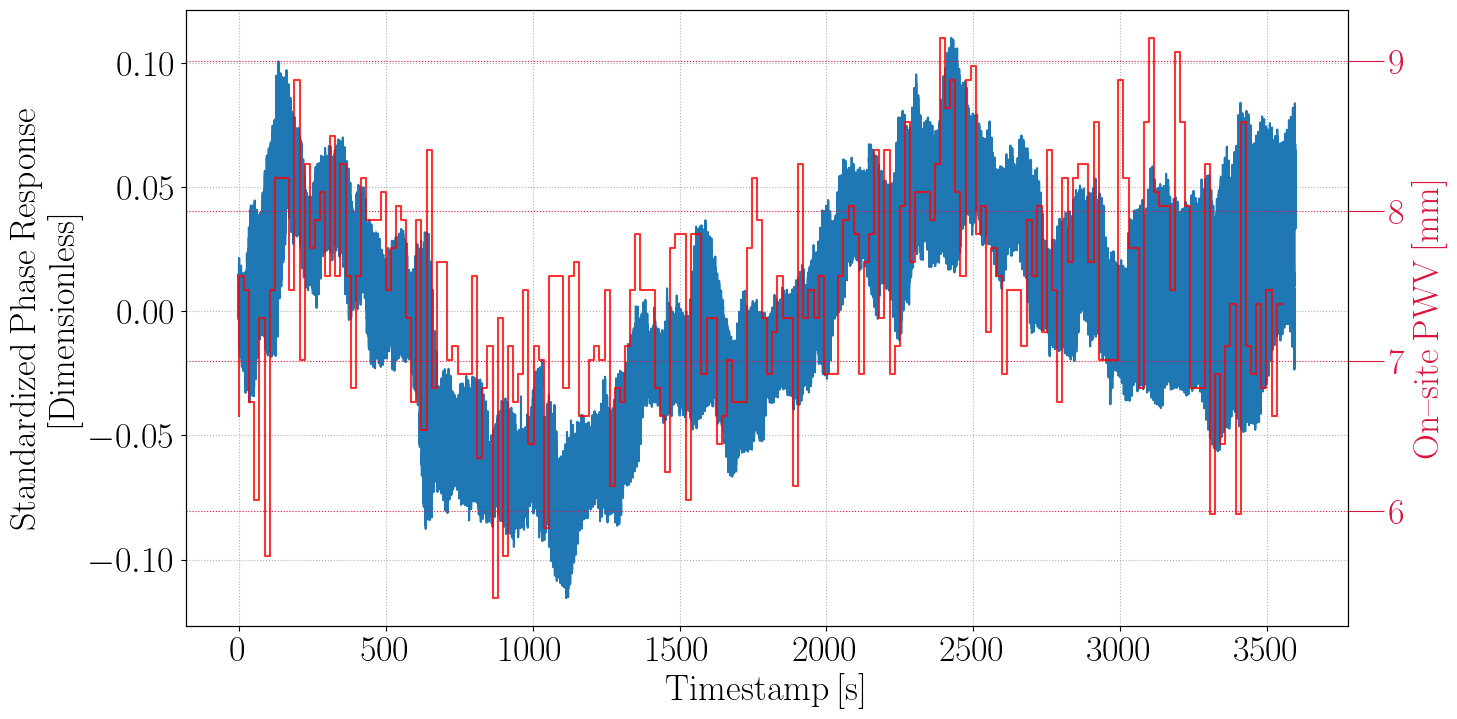}
}
\hspace{1pt}
\subfloat[]{
\includegraphics[width=0.985\columnwidth]{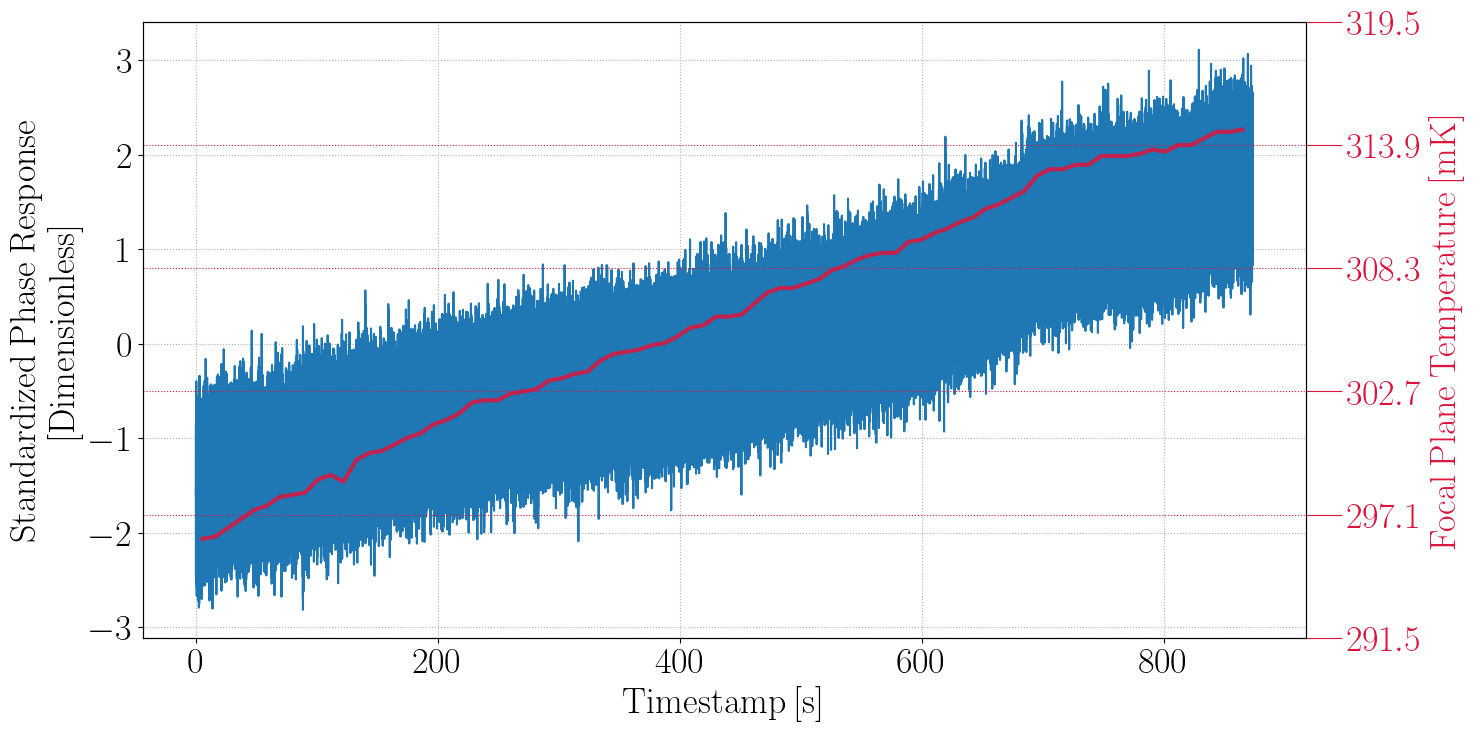}
}
\caption{Standardized phase response TOD plots, measured under different conditions. In (a), one hour TOD from $\textrm{MKID\:1}$ on $\textrm{Array2B}$ (blue, left axis) measured at 9 RPM and evolution of the precipitable water vapor (red, right axis) are shown. The TOD shown in (b) was measured by $\textrm{MKID\:0}$ on $\textrm{Array2A}$ while intentionally varying the focal plane temperature conditions. The red curve indicates the focal plane temperature measured during a 20 RPM scan, prior to reaching thermal stabilization.}
\label{fig:TODs}
\end{figure*}

\section{MKID Systematics Due to Background Evolution and Thermal Instability}

Although laboratory measurements verified the suitability of the prototype MKIDs for scientific observations \cite{kutsuma-phd}, the on-site data revealed significant contributions from both atmospheric emission and the focal plane temperature variations. The primary features of long-term drift and fluctuations observed in TODs were found to be correlated with variations in the PWV and focal plane temperature, as shown in Fig. \ref{fig:TODs}. The phase responses shown here have been standardized by dividing by their standard deviations. In (a), one hour phase response from $\textrm{MKID\:1}$ on $\textrm{Array2B}$ is shown in blue, overlaid with the PWV measurements in red. The blue curve in (b) shows a $\textrm{10\:}$minute long TOD from $\textrm{MKID\:0}$ on $\textrm{Array2A}$, measured during a controlled variation of the focal plane temperature. The overlaid red curve shows the focal plane temperature measured at a rotation speed of 20 RPM, prior to reaching thermal equilibrium. In this section, we present characterizations of the MKID response to these two sources.

\subsection{Characterizing Atmosphere-Induced Systematics}

The atmospheric water vapor content is the dominant contributor to in-band variations in background emission for millimeter and sub-millimeter wavelengths \cite{Errard_2015}. The absorption of incoming photons impacts the quasiparticle population in the MKIDs, leading to measurable shifts in resonance frequency. Since it is the relative frequency shifts that were analyzed, constant background contributions such as those from the telescope's mirror system or the optical filters can be ignored in these measurements. In this study, atmospheric variations were tracked through the PWV, which quantifies the water content in the atmosphere.

\begin{figure}[t]
  \centering
  \includegraphics
  [width=0.98\columnwidth]{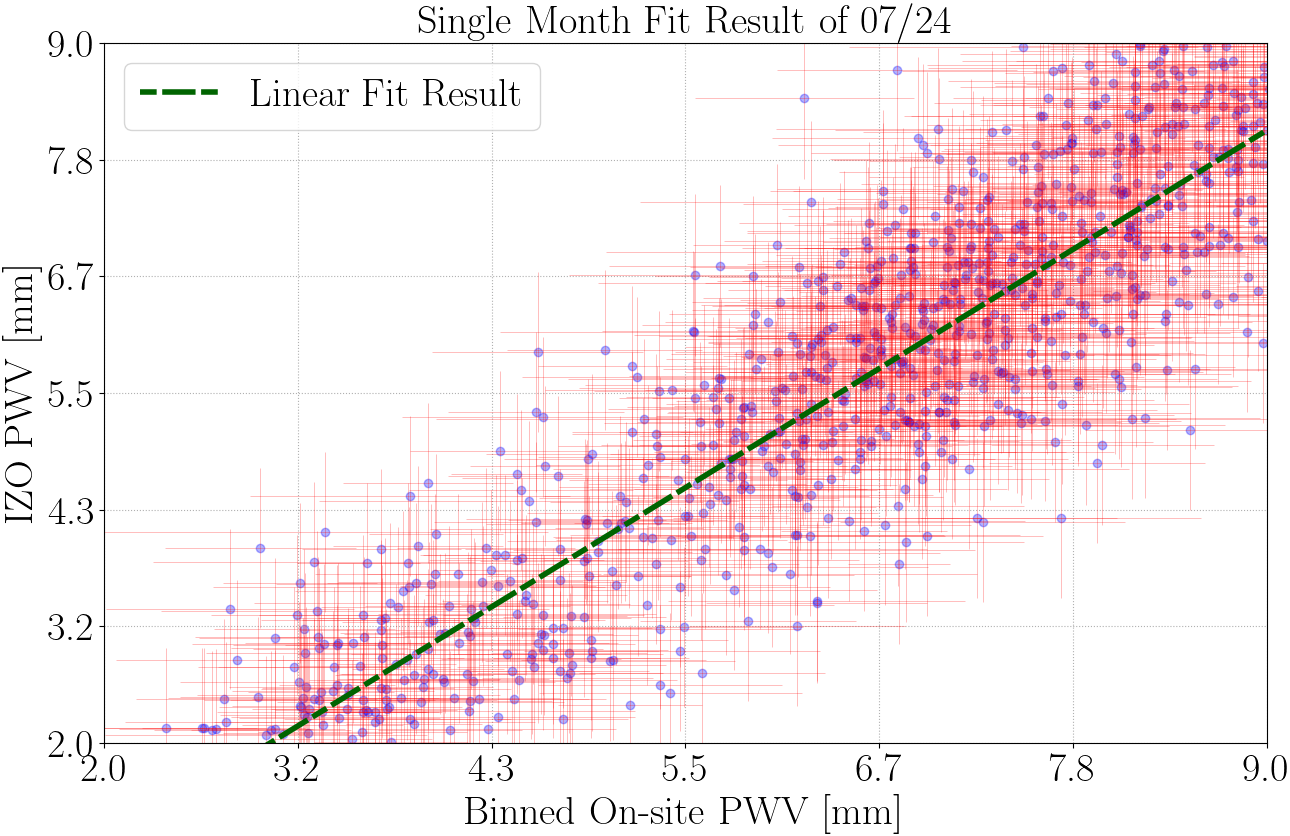}
  \caption{Plot of the Izaña atmospheric observatory} PWV data against $3$-minute binned on-site PWV measurements from July of 2024. The uncertainty bars depict standard deviations in the measured PWV. The green dashed line shows the linear fit result.
  \label{fig:monthly pwv}
\end{figure}

We utilized two PWV databases. Instituto de Astrofisica de Canarias (IAC)'s Izaña atmospheric observatory (IZO) measurements \cite{julio_WRF}\cite{gaulli-url}, and the on-site radiometer measurements. The IZO database has been validated through its extensive data based on the Global Navigation Satellite System, but it produces just two measurements every hour. The on-site radiometer produces 3 measurements per minute by scanning the 23 GHz water vapor line. Its relatively fast sampling speed enables us to capture short-term PWV variations. However, it has not yet been adopted in astronomical observation data corrections. The following analysis serves as a preliminary calibration study.

The on-site PWV measurements are corrected based on the IZO data through monthly linear fits, as shown in Fig. \ref{fig:monthly pwv}. The $y$ axis uncertainty bars are taken from the IZO database, while the $x$ axis uncertainty bars represent the standard deviation of $3$-minute binned on-site measurements. 

To model the dependence of the MKID resonance frequency $f_r$ on atmospheric emissions parametrized by the PWV, the band-averaged line of sight atmospheric radiation power incident on the detector $\mathit{P_{rad}}$ at atmospheric temperature $\mathit{T_{atm}}$ can be derived and approximated based on \cite{Essential-Radio-Astronomy} and simplified as \begin{align}
\label{eq:P_atm}
P_{rad}(PWV)&=P_{max}(\langle T_{atm}\rangle )\cdot(1-e^{-\tau \cdot m})\:. 
\end{align}
\vspace{0.1mm}
In this formulation, the atmospheric temperature is approximated by a time-averaged value over the observation periods, represented as $\langle\mathit{T_{atm}}\rangle$, in kelvin. $\mathit{P_{max}}$ is the black-body approximation of the atmospheric radiation power incident on the focal plane in watts, integrated over the bandpass and the optical efficiency. $\tau$ is the dimensionless band-averaged atmospheric opacity along the zenith, and the airmass $\mathit{m}$ is a dimensionless value approximated as the secant of the telescope's zenith angle and is considered constant throughout the data set, as all the observations were conducted at a fixed elevation of $70^{\circ}$.

Second, the MKID properties are governed by the quasiparticle density inside it. The total density of the quasiparticle $\mathit{n^{qp}_{tot}}$ can be approximated as the sum of the quasiparticle density generated from the optical signal $\mathit{n^{qp}_{optical}}$, and the thermally generated quasiparticle density $\mathit{n^{qp}_{thermal}}$ \cite{kutsuma-phd}, given as 
\begin{eqnarray}
n^{qp}_{tot} &=& n^{qp}_{optical}+n^{qp}_{thermal}\:. 
\label{eq:n^tot_qp}
\end{eqnarray}

As with other typical MKID systems \cite{gao-phd}, GroundBIRD is assumed to be optically dominated, and hence only the optical quasiparticle density is taken into account for this analysis. Using equation 7.3 in \cite{mazin-phd}, the optical quasiparticle density can be expressed as 
\begin{eqnarray}
n^{qp}_{tot}&\approx&n^{qp}_{optical}=\frac{\eta_{pb}\tau_{qp}}{V\Delta}\eta_{opt}P_{rad}\:,
\label{eq:n^opt_qp}
\end{eqnarray}

\noindent where $\mathit{\eta_{pb}}$ is the detector's pair-breaking efficiency, $\tau_{qp}$ is the quasiparticle lifetime, $V$ is superconducting volume of the detector, $\Delta$ is the superconducting energy gap, and $\eta_{opt}$ is the GroundBIRD optical efficiency of $0.55$ \cite{tomonaga-phd}. 

Following equation 2 in \cite{catto} for the kinetic inductance $L_{ki}$, assuming an ideal superconductor, and taking the first-order expansion about $n^{qp}_{tot}$ gives 
\begin{eqnarray}
\label{eq:Lk}
\frac{\delta L_{ki}}{L_{ki0}}  &\approx& \frac{\delta n^{qp}_{tot}}{n_{s0}}.
\end{eqnarray}

\noindent $L_{ki0}$ is kinetic inductance at T = 0 K and $n_{s0}$ is twice the Cooper pair density at T = 0 K.

The fractional shift in the MKID resonance frequency $f_r$ as a function of kinetic inductance is given by Eq. \ref{eq:fr vs Lk} where $\alpha$ is the kinetic inductance fraction and $f_{r0}$ is the reference resonance frequency \cite{gao-phd}. Then inserting Eq. \ref{eq:n^opt_qp} yields a relation between the fractional shift of the resonance frequency and shift in the relative quasiparticle density, described as
\begin{eqnarray}
\frac{\delta f_r}{f_{r0}}&=&-\frac{\alpha}{2}\frac{\delta L_{ki}}{L_{ki}} \label{eq:fr vs Lk} \\
&\approx&-\frac{\alpha}{2}\frac{\delta n^{qp}_{tot}}{n_{s0}}\:.
\label{eq:fr vs nqp}
\end{eqnarray}

Finally, applying a linear approximation for the optical depth as a function of PWV \cite{de-gregori} yields a three-parameter model of the MKID resonance frequency as a function of PWV, given as
\begin{equation}
    \begin{split}
        &f_r(PWV)=A+B\cdot e^{-C\cdot PWV} \label{eq:fr vs pwv}\:.
    \end{split}
\end{equation}

\begin{figure*}[!t]
\centering
\subfloat[]{
\includegraphics[width=0.985\columnwidth]{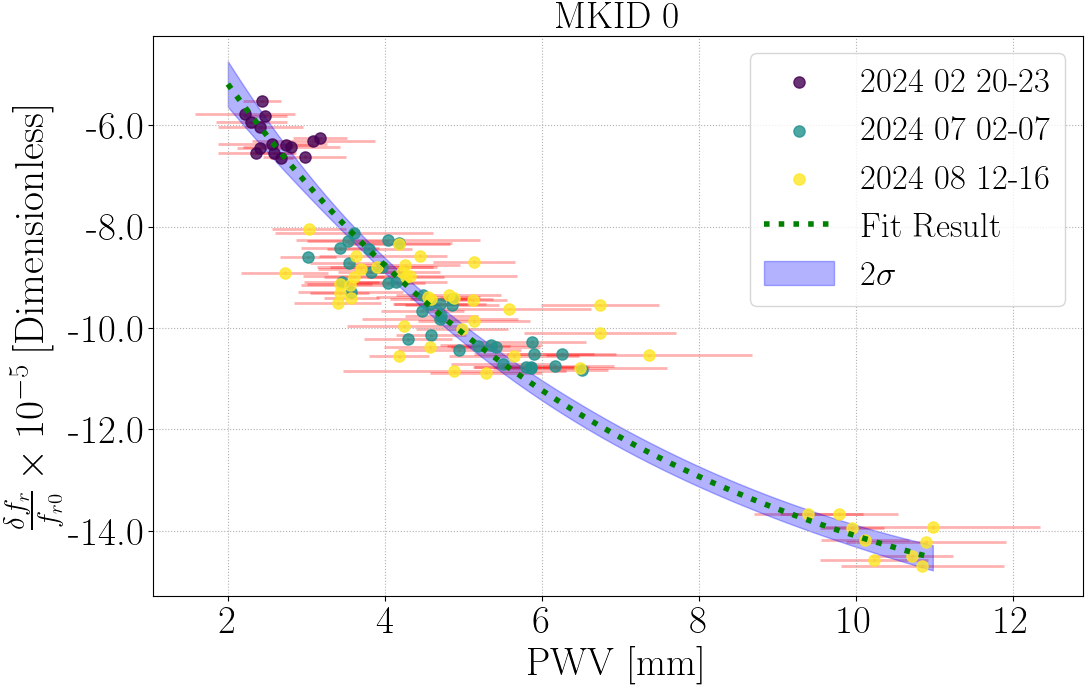}
}
\hspace{1pt}
\subfloat[]{
\includegraphics[width=0.985\columnwidth]{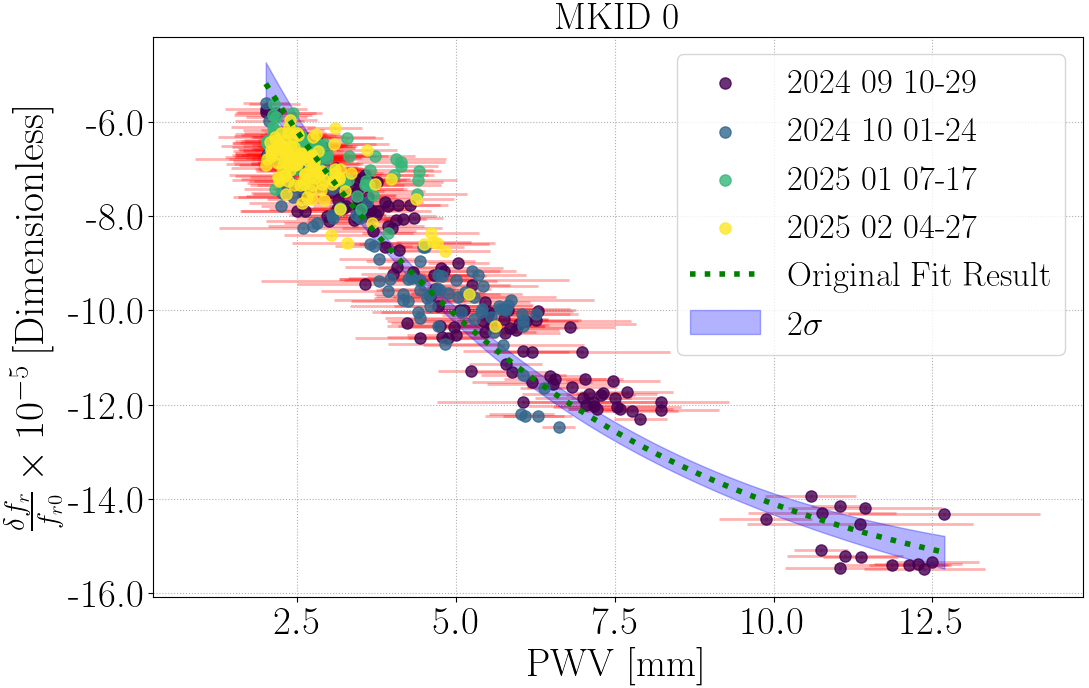}
}
\caption{Fractional resonance frequency shift $\frac{\delta f_r}{f_{r0}}$ of \textrm{MKID 0} on \textrm{Array1B} plotted against calibrated on-site PWV. The reference resonance frequency $f_{r0}$ is taken as the model-estimated value at $0\:\mathrm{mm}$ PWV. Colored points indicate different observation periods. The plot in (a) shows the primary dataset with the best-fit model (green dotted line) and $\pm2\sigma$ confidence interval (blue shaded region). The plot in (b) displays the same fit and confidence interval overlaid on an independent dataset for validation.}
\label{fig:fr vs pwv data fit}
\end{figure*}

\noindent Building on the model presented above, the measured dependence of the fractional shift in MKID resonance frequency on PWV is characterized. The model-estimated resonance frequency at PWV $=0\:\mathrm{mm}$ is taken as the reference resonance frequency $f_{r0}$. Fig. \ref{fig:fr vs pwv data fit} (a) presents the plot of fractional resonance frequency shift against PWV with data from different observation periods shown in different colors. The green dotted line shows the best-fit model, with $\pm2\sigma$ confidence interval shaded in blue. Uncertainties derived from the resonance peak fits with the sweep measurements are taken as uncertainty bars for the $y$ axis, although they are sufficiently small that they do not appear in the figure. On-site PWV measurements at 3 samples per minute were aggregated into $3$ minute bins, with the standard deviations of each bin assigned as the $x$ axis uncertainty.

The fits were independently carried out on measurements from multiple detectors, and their reduced-$\chi^2\:(\chi^2_{red})$ values are summarized in Table \ref{tab:fit results} (Dataset 1). A median $\chi^2_{red}$ value of $1.27$ provides a statistically consistent fit to the resonance frequency shifts as a function of PWV. Slight deviation from unity suggests possible unmodeled contributions, including the use of a time-averaged approximation for the atmospheric temperature noted above. The array-averaged values of the exponential parameters are given as $\mathrm{B}=(1.5\pm0.11)\times10^{-4}$ and $\mathrm{C}=(2.0\pm0.19)\times10^{-1}$.

The original fit parameters were applied to an independent dataset. Fig.\ref{fig:fr vs pwv data fit} (b) shows the fractional resonance frequency shift against PWV of this new dataset, with the original fit result overlaid. The $\chi^2_{red}$ values, calculated using the fixed parameters on the new dataset, are listed for each detector in \ref{tab:fit results} (Dataset 2). With a median $\chi^2_{red}$ value of $1.34$ on the independent dataset using the original fit result, the model demonstrates a reliable description of the resonance frequency shift as a function of PWV.

\renewcommand{\arraystretch}{1.45}
\begin{table}[!t]
\caption{$\chi^2_{red}$ Values for the Resonance Frequency vs. PWV Fits from MKIDs of $\textrm{Array1B}$} 
\label{tab:fit results}
\centering
\begin{tabular}{|c|c|c|}
\toprule
       & Dataset 1 & Dataset 2\\
\hline
 MKID 0 & 1.30 & 1.19\\
 \hline
 MKID 1 & 1.19 & 1.34\\
 \hline
 MKID 5 & 1.23 & 1.19\\
 \hline
 MKID 6 & 1.27 & 1.21\\
 \hline
 MKID 7 & 1.15 & 1.19\\
 \hline
 MKID 8 & 1.27 & 1.19\\
 \hline
 MKID 11 & 1.32 & 1.47\\
 \hline
 MKID 12 & 1.20 & 1.23\\
 \hline
 MKID 13 & 1.28 & 1.56\\
 \hline
 MKID 14 & 1.33 & 1.74\\
 \hline
 MKID 15 & 1.37 & 1.43\\
 \hline
 MKID 20 & 1.35 & 1.37\\
 \hline
\hline
Array Median & 1.27 & 1.34\\
\bottomrule
\end{tabular}
\end{table}

\subsection{Thermal Contributions on the Systematic Effects}

Although optical loading typically dominates in MKID telescopes, the effect from the thermal drifts of the focal plane must be evaluated to achieve higher performances, as their slow variations can impact the large-scale CMB signals targeted by this experiment.

\begin{figure}[t]
  \centering
    \includegraphics[width=0.99\columnwidth]{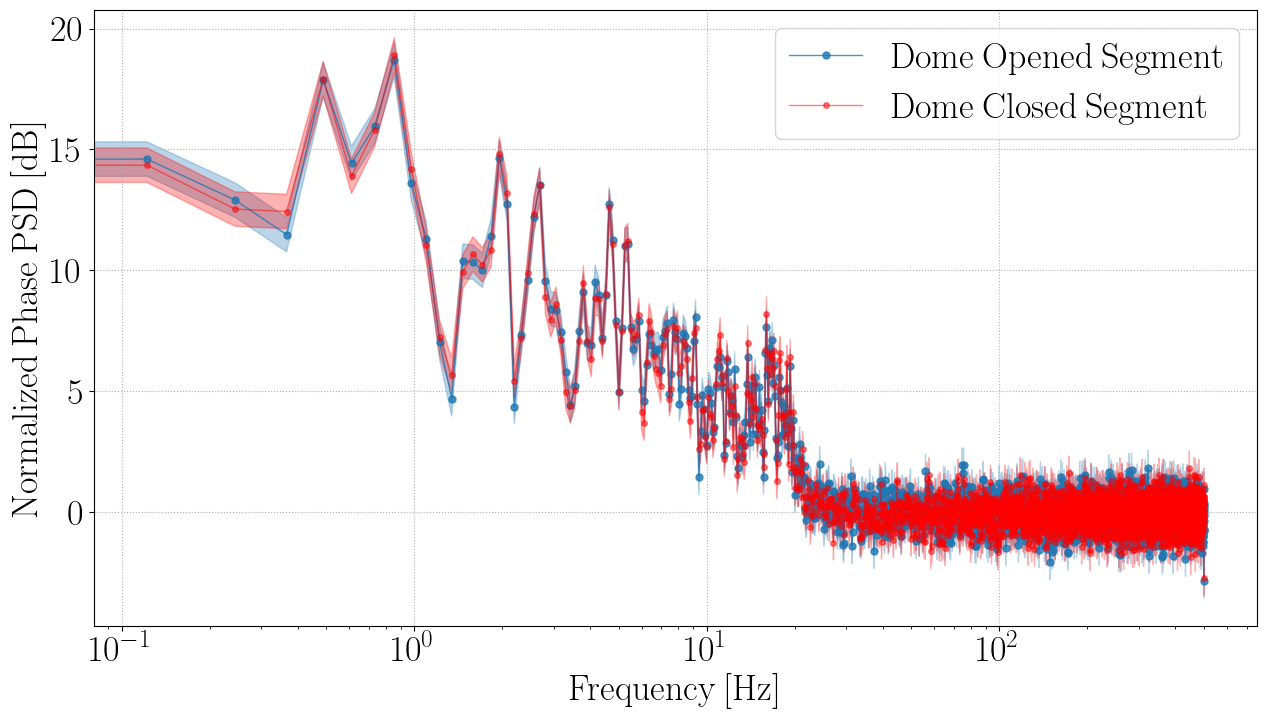}
  \caption{PSDs of a dark MKID phase response for dome-open (blue) and dome-closed (red) segments, normalized to the high-frequency white-noise level of the open-dome data. Each shaded region show $\mathrm{\pm1\sigma}$ uncertainty from Welch's method. The RMS difference is 0.84 times the mean effective uncertainty.}
  \label{fig:dome_psd}
\end{figure}

Measurement with dark MKIDs were performed to investigate non-optical systematics affecting the detectors. Their insensitivity to optical load was confirmed through observation of bright thermal source, such as the protective dome exposed to the external environment. Fig. \ref{fig:dome_psd} compares the power spectral densities (PSD) of the phase response measured during a dark MKID observation. The first segment of the observation was conducted with the protective dome open (blue) followed by a second segment with the dome closed (red). Both spectra were normalized to the mean power in the high-frequency white-noise tail of the dome opened data. Shaded regions indicate $\mathrm{\pm 1\sigma}$ statistical uncertainty on each PSD, derived from the effective number of independent segments in Welch's method. The root mean squared (RMS) difference between the spectra is 0.84 times the mean uncertainty, indicating no statistically significant difference and confirming the detector's insensitivity to an external source.

\begin{figure}[!t]
  \centering
  {
    \includegraphics[width=0.98\columnwidth]{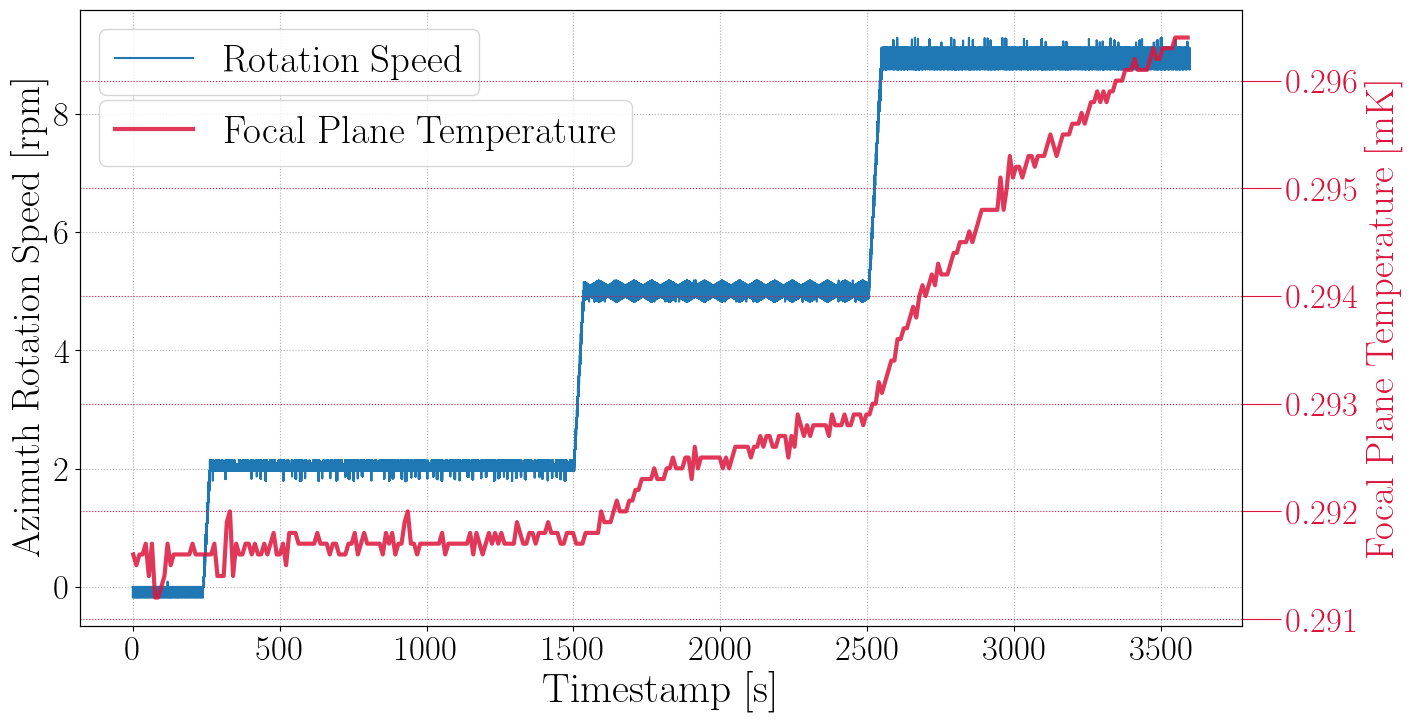}
  }
  \caption{Telescope azimuthal rotation speed (blue, left axis) and focal plane temperature (red, right axis) during discrete increases in rotation speed, with the protective dome open. The rotation speed reached up to $12$ RPM, due to difficulties managing thermal stability of the focal plane at higher speed.}
  \label{fig:rotation}
\end{figure}

Due to the absence of an active control system of the temperature at the focal plane, its temperature was modulated indirectly by increasing the telescope's azimuthal rotation speed. We took advantage of the transient time window following each speed change, during which the focal plane temperature had not yet stabilized, to study its effect on the resonance frequency. Fig. \ref{fig:rotation} shows increases in the focal plane temperature during these transient periods following the changes in rotation speed. In addition to rotation-induced heating, we also observed a small temperature increase of less than 1mK when first switching on the readout system, which is excluded in this analysis.\\

Since the effect on dark MKIDs from the shifting optical load is minimal, the optically generated quasiparticle density $n^{qp}_{opt}$ can be ignored. The resulting total quasiparticle density is given as a function of the focal plane temperature $T$ \cite{visser2011}, described as
\begin{align}
n^{qp}_{th} &\simeq 2N_0 \sqrt{2\pi \Delta k_B T}\cdot e^{-\Delta/k_B T}\:.
\label{eq:n^th_qp}
\end{align}

\noindent $N_0$ is the detector's electron density of states at the Fermi level, and $k_B$ is the Boltzmann constant. The Bardeen-Cooper-Schrieffer (BCS) theory states that $\Delta$ at temperatures below the superconducting transition temperature $T_c$ can be approximated and expressed as \cite{bcs} 
\begin{align}
2\Delta &=3.52k_B T_c\:.
\label{eq:Delta}
\end{align}

\noindent To account for the offset between thermometer and detector locations, we replace the measured temperature $T$ with an effective value $T + \Delta T$, where the offset $\Delta T$ is treated as an additional free parameter fitted separately for each detector. Inserting Eq. \ref{eq:Delta} into Eq. \ref{eq:fr vs nqp} and replacing $T$ to $T+\Delta T$ yields a resonance frequency model as a function of the focal plane temperature, described as
\begin{align}
f_r(T)&=A\cdot \sqrt{(T+\Delta T)}\cdot e^{-B/(T+\Delta T)}+C\:.
\label{eq:fr vs T model}
\end{align}

\noindent In this expression, the exponential scale factor $B=\Delta/k_B$, is taken as a physical constant under our operating conditions. With $T_c\approx1.2$ K for bulk aluminum\cite{al_Tc}, $B$ is fixed to $2.2$ K, resulting in a three-parameter model with only $A$, $C$ and $\Delta T$ as free parameters.

The temperature is monitored using a calibrated Lake Shore RX-202A ruthenium oxide sensor. The calibration specification sheet provides a standard uncertainty of $\pm2\:\mathrm{mK} \:(1\sigma)$ at our operating temperature ($\sim 290\:\mathrm{mK}$) \cite{LakeShore}.

\begin{figure*}[!t]
\centering
\subfloat[]{
\includegraphics[width=0.985\columnwidth]{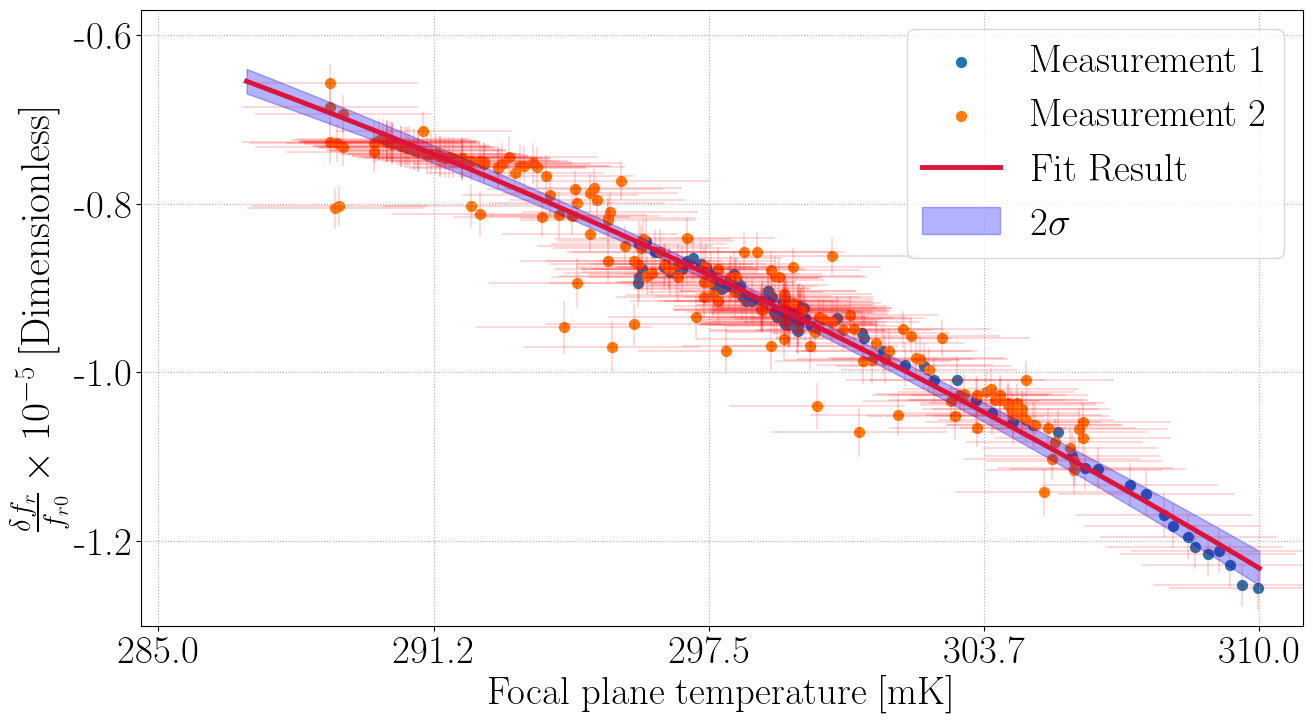}
}
\hspace{1pt}
\subfloat[]{
\includegraphics[width=0.985\columnwidth]{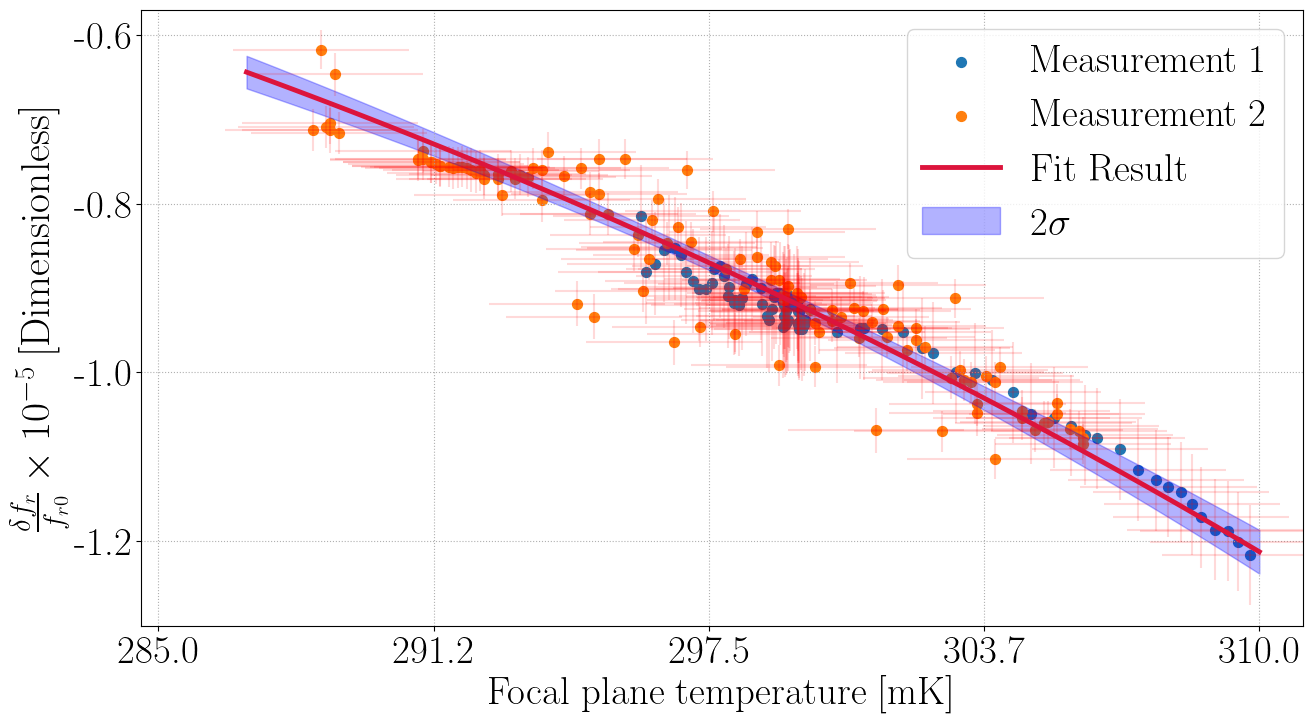}
}
\caption{Fractional resonance frequency shift as a function of the focal plane temperature from two different dark MKIDs. Blue and orange points represent data from separate observation periods. The thick red line is the best-fit model, with the blue shaded region showing the $\pm2\sigma$ confidence interval. The plot in (a) shows results taken by a detector in \textrm{Array3B} and the plot in (b) shows results from a detector in \textrm{Array1B}.}
\label{fig:fr vs T}
\end{figure*}

Fig. \ref{fig:fr vs T} shows the fractional shift in the resonance frequency of dark MKIDs against the focal plane temperature, obtained from different detectors, with the thick red curve representing the best fit result, and blue shaded region showing the $\pm2\sigma$ confidence interval. The limiting resonance frequency of the model at $\textrm{T}\xrightarrow{}0\:\textrm{K}$ is taken as the reference resonance frequency.

The data align with the model within the $2\sigma$ confidence interval across the nominal operating range (generally below $300\:\mathrm{mK}$ for aluminum; we extended to $310\:\mathrm{mK}$), supported by the ensemble-averaged $\chi^2_{red}$ of $0.3$ across 6 dark MKIDs distributed over the focal plane. The average fitted thermal offset $\langle\Delta T\rangle=(0.3\pm0.2)\:\mathrm{mK}$, and the scaling parameter $\mathrm{\langle A\rangle=(3.3\pm0.6)\times10^{-2}}$.

We extended measurements to temperatures beyond 310 mK to assess the model's robustness at elevated temperatures. Although the measurements are still statistically compatible with the model at the $2\sigma$ level, a systematic deviation becomes apparent at higher temperatures. Several effects likely contribute to this behavior. As temperature increases, enhanced quasiparticle density and scattering degrade the precision and reliability of resonance frequency determination. 
Additionally, during rapid thermal excursions induced by high-speed rotation, the thermometer reading may not accurately track instantaneous detector temperatures, potentially introducing systematic bias.

\subsection{Relative Contribution of Atmospheric and Thermal Effects}
Relative effects of the two systematic sources are compared under typical observation conditions. The average on-site PWV from July 2023 to August 2024 was 4.3 mm, with a typical hourly fluctuation of $0.7\:\mathrm{mm}$, calculated as the standard deviation within 1-hour observation. Using the atmospheric model, this variation results in an average fractional frequency deviation of $\sim 10^{-5}$ over one-hour timescale. In comparison, the focal plane temperature averages $288.9\:\mathrm{mK}$ during typical observations, with an hourly deviation of $0.2\:\mathrm{mK}$, leading to a mean modeled fractional frequency deviation of $\sim10^{-7}$. These estimations indicate that resonance frequency fluctuations in a typical hour-long observation due to PWV variation are more than two orders of magnitude larger than those caused by focal plane temperature changes, confirming atmospheric emission as the dominant source of environmental systematics in our setup. This puts emphasis on atmospheric condition in both observing strategy and data processing pipeline, while thermal variations remain important for long-timescale drifts.

\section{Conclusion}
This study examined the systematic variations identified in the resonance frequency shifts of the MKID employed in the GroundBIRD telescope.

The frequency response induced by optical loading, quantified in terms of atmospheric PWV. Since the on-site PWV monitor provides high time resolution but is subject to calibration uncertainty, it was corrected using data from the IAC's IZO measurements. The resulting model connects atmospheric PWV to the in-band optical loading through the system's bandpass filters and ultimately to the MKID frequency response. When applied to data from multiple detectors, the model achieves a mean $\chi^2_{red}$ of $1.27$. The model maintains comparable statistical quality $\chi^2_{red}$ of $1.34$ when applied to an independent dataset, validating its predictive capability across different observing periods.

Thermal effects were investigated using dark MKIDs, allowing us to isolate the effect of focal plane temperature variations on the frequency shift. The thermal model is grounded in the thermal quasiparticle density derived from the BCS theory, and yielded an averaged $\chi^2_{red}$ of $0.3$ across 6 dark MKIDs, showing the model effectively describes MKID's frequency response to thermal shifts under nominal operating focal plane temperature.

Atmospheric emission dominates over thermal effects under typical observation conditions by more than two orders of magnitude, establishing the priority for PWV-based corrections in the observation and data analysis pipeline. Nevertheless, achieving thermal stabilization remains critical to prevent the signal to be affected by systematics, as shown in this study.

The results presented here point to several improvements and future studies. A correlation was found between the focal plane temperature and the azimuth rotation, with stable operating temperature maintained up to 9 RPM in the current configuration. Future maintenance work aims to further improve thermal stability and achieve lower bath temperature during faster azimuth scanning. In addition, the current fixed one-hour duration used for both TOD measurement and resonance frequency sweep measurement interval could be made to change dynamically according to the PWV. This would allow the system to follow the evolving atmospheric emission more effectively, reducing the impact of time-dependent optical systematics on the detector response in each observation cycle.

\section*{Acknowledgments}
This article was partially supported by the Korea University Research Grant and the Research Support Grant RS-2022-NR068913. We are partially supported by high-speed KREONET provided by KISTI. The on-site PWV measurements were taken with a radiometer manufactured by Furuno Electric Co., Ltd.

\bibliographystyle{IEEEtran}
\bibliography{LTD2025_ref}

\end{document}